# Sustainable allocation of greenhouse gas emission permits for firms with Leontief technologies

E. Gutiérrez, N. Llorca, J. Sánchez-Soriano*
CIO and Department of Statistics, Mathematics and Computer Science
University Miguel Hernández of Elche, Spain
M. Mosquera
Department of Statistics and Operations Research
University of Vigo, Spain

**Abstract**

In this paper we deal with production situations where a cap or limit to the amount of greenhouse gas emissions permitted is imposed. Fixing a tax for each ton of pollutant emitted is also considered. We use bankruptcy rules to define cooperative games with externalities associated with these situations and analyze the existence of coalitionally stable allocations of the emission permits. We prove that the constrained equal awards ($CEA$) rule provides stable allocations and as a direct mechanism, it is incentive compatible. These two facts have interesting managerial implications to control pollution emissions.

## 1 Introduction

Concern about climate change and, in particular, about global warming in the atmosphere is nothing new. In 1992, the Framework Convention on Climate Change (FCCC) took place in Rio de Janeiro (UNFCCC, 1992), in which the signatory countries pledged to take measures to avoid climate change, but without setting out specific measures. In 1997, the FCCC took place in Kyoto, from which the so-called Kyoto Protocol (UNFCCC, 1998) came into being, whereby the signatory countries committed themselves to reducing their greenhouse gas

*Corresponding author, e-mail: joaquin@umh.es

(GHG) emissions to certain levels until 2020 (considering the Doha Amendment to the Kyoto Protocol in 2012), but did not establish the procedure that each country should follow to achieve its emission target. More recently, in 2015, the FCCC took place in Paris, from which the so-called Paris Agreement (UNFCCC, 2015) came into being. In this binding agreement 188 countries have committed to controlling their GHG emissions, contrary to the Kyoto Protocol where only certain countries committed, and have indicated their national contributions will be subject to a gradual reduction every five years, see Carraro (2016) for an interesting summary. Thus, each country has a limit or target for each period to be divided among the sectors involved.

The most common approaches in economic theory to control pollution emissions involve the use of taxes and cap-and-trade systems. Tax systems are price instruments in which the government agency fixes the price per unit of emissions (tax), but the quantity remains unknown. In contrast, cap-and-trade systems are quantity instruments in which the authority fixes the quantity of emissions allowed (cap), but the price per unit is determined by means of a certain market (trade). The European Union Emissions Trading Scheme is probably the best known trading scheme. Nevertheless, both systems are widely used in practice, see Carl and Fedor (2016) for a survey of the carbon revenues from the cap-and-trade and carbon tax systems in the World, together with uses of those revenues by the governments. These systems have attracted considerable attention for many years and there is a large volume of literature. There are many papers comparing the efficiency, advantages and disadvantages of both systems. For example, Cooper (1998), Nordhaus (2007) and Avi-Yonah and Uhlmann (2009) suggest that carbon taxes are better than cap-and-trade systems. However, Keohane (2012) suggests that cap-and-trade systems have interesting advantages when compared with the application of taxes. A different option is the so-called safety valve, Jacoby and Ellerman (2004), which is somewhat of a combination of both methods.

This paper falls within the latter idea of combining quantity instruments (cap-and-trade) and price instruments (carbon taxes). Therefore we try with our model to contribute by giving an insight into how to mitigate some of the potential/possible shortcomings of both models: cap-and-trade and tax systems. In particular, the overestimation of the limit in the cap-and-trade system, the no-control of the abatement of emissions, if any, in the tax system, and the difficulties of measuring how much companies actually pollute in both systems. The first two drawbacks are related to the possibility of lack of success in the abatement of emissions and the latter is associated with the authority's knowledge of the real emissions.

The economic and social implications that arise from the previous agreements and control instruments can be studied from different fields of OR/MS, depending on the analysis that is intended to be carried out. See Tang and Zhou (2012) for a survey on OR/MS research developments in enviromental and sustainable issues; and Finus (2001), Dinar et al. (2008) and Patrone et al. (2008) for applications of Game Theory to natural resources management and environmental problems. Some interesting problems that arise are the allocation and management of GHG emission permits or allowances; the effect of GHG emission control systems on the behavior of companies; the analysis of international agreements; the effects of these measures on competition, collaboration or co-opetition between companies, etc.

In this paper we study the problem of allocation of carbon dioxide emission permits for firms by using bankruptcy rules. We analyze the consequences in the cooperation among companies when we consider that externalities can arise. In order to do so, we use the context production situations in which there is a cap on the emissions, ex ante cooperation among companies by utility transfers is possible, trading among companies is allowed and there is a fixed tax on the carbon dioxide consumption. To the best of our knowledge, no research has examined this approach to the allocation of GHG emission permits. Our research aims to fill this gap in the literature by examining the following key questions:

a) How to model externalities when companies can coordinate their claims on carbon dioxide emission permits and the allocation of those permits is to be carried out by using a bankrutpcy rule?

b) How to allocate carbon dioxide emission permits to obtain a stable allocation of the global profits among the companies?

c) How the paremeters defining the problem (cap, allocation rule and tax) can be fixed or used in order to manage and control the carbon dioxide emissions efficiently?

To this end, we consider production situations where several firms own resource bundles that can be used to produce various products which they sell at the given market prices. All firms have the same production function but differ in the amount of resources which they can manage, so they are different in size. Under the market conditions mentioned above, using games with externalities or in partition function form (Thrall and Lucas, 1963) together with arbitration by applying bankruptcy or rationing techniques (O'Neill, 1982; Aumann and Maschler, 1985; Curiel et al., 1987), we examine the process of allocating the carbon dioxide emission permits in order to analyze under what conditions we can obtain stable allocations of the emission permits and stable distributions of the total profit generated by the market. Likewise, we study how the cap on the emissions, the allocation rule and the per unit tax on the emissions can be used for managing and controlling the gas emissions.

The rest of the paper is organized as follows. In Section 2 a review of the related literature is presented. Section 3 describes our proposal. In Section 4 some preliminaries on TU-games and bankruptcy problems are given together with the description of linear production situations with an external limited resource ($LPP$ situations). In Section 5 a further approach is provided by using bankruptcy techniques to deal with $f-LPP$ situations and the $f-LPP$ games with externalities associated with these situations and we prove that if there are stable allocations of the permits, then there are stable allocation of the total profits. In Section 6, if the total demand exceeds the cap, we prove that, under certain conditions, using the $CEA$ rule the allocation of carbon dioxide emission permits obtained is coalitionally stable, i.e., there is no group of firms that can complain by arguing it is unfair. Moreover, we show that the $CEA$ rule as a direct mechanism is incentive compatible. In Section 7 we describe the managerial implications of our proposal. Some concluding remarks are given in Section 8. All proofs of the results may be found in the appendix.

# 2 Literature review

In this section we review the literature of two different topics related to the abatement of GHG emissions: (i) the impact of GHG emission control policies on the behavior of the companies with respect to their operational decisions, (ii) how to allocate carbon dioxide emission permits. Finally, we review the more specific game-theoretic literature related to our model.

The effects of the GHG emission control policies on the operational decisions of the companies have been studied during recent years from OR/MS perspective. Some recent papers are Bai and Chen (2016) and Hong et al. (2016), where there is only one decision maker: the firm. However, the analysis of the behavior of the companies when there is more than one decision maker is usually carried out by using some game-theoretic approach. Among the latest works, He et al. (2012) compares the effectiveness and efficiency of cap-and-trade and tax systems in the context of power generation in terms of different criteria. For their analysis, they use a Nash-Cournot competition model. In the particular case of the cap-and-trade systems, the agents must purchase the allowances of emissions at an exogenously given price. The grandfathering case for the cap-and-trade case is studied in He et al. (2010). Luo et al. (2016) consider two competing manufacturers who have different emission reduction efficiencies and study the effects of pure competition and co-opetition on emission reduction efficiency under a cap-and-trade policy. Yenipazarli (2016) uses a multi-stage leader (regulator)-follower (firm) Stackelberg game model to investigate the impact of emission taxes and emissions trading on the optimal production (new product or remanufactured) and pricing decisions of a manufacturer. Siddiqui et al. (2016) study the impact of market structure with renewable portfolio targets and show that social welfare under perfect competition between renewable and non-renewable is lower than when the non-renewable energy sector exercises market power in a Cournot oligopoly. However, we have not found any paper that studies the effect of the externalities that can arise when an allocation mechanism is set to distribute the carbon dioxide emission permits, if the coordination and prior compensation among companies is allowed when requesting the permits.

With regard to the GHG emissions allocation, Zhou and Wang (2016) review the literature on the carbon dioxide emission allocation and classify the allocation methods into four groups: indicator, optimization, game-theoretic and hybrid approaches. Moreover, they distinguish between different application levels: countries, regions or firms. Likewise, they conclude that the game-theoretic allocation methods are based on cooperative games, dynamic games and incomplete information games. However, bankruptcy techniques are not mentioned in this survey. Giménez-Gómez et al. (2016) propose the use of bankruptcy rules, based on the selection of some desirable principles, as mechanisms to allocate the global carbon budget among countries. Kampas and White (2003) examine a variety of permit allocation rules in order to allocate nitrate emissions for a small catchment in South West England. Some of the permit allocation rules that they study are bankruptcy rules. They then compare the results obtained by applying the allocation rules and the results obtained from an asymmetric Nash's bargaining solution in order to compare the correspondence between ex-ante and ex-post criteria of equity. Nevertheless, the above mentioned papers mainly focus on the direct application of the bankruptcy rules, but they do not embed the allocation rule inside the game in order to consider the possible externalities arising from the process of allocation and then study the stability of the final distribution of the emission permits and the final distribution of the total revenue of the whole system. So we analyze not only the impact of a control policy in the behavior of companies with regard to their operational decisions, but also the effect of the allocation rule in their strategic behavior for obtaining an advantageous allocation of permits.

In this paper we use a basic production model, the so-called linear production (LP) model since the programs that arise are linear. These situations and the corresponding cooperative games were introduced by Owen (1975). Some extensions of this model were introduced by Molina and Tejada (2006) and Tijs et al. (2001). Recently, linear production situations in which there is a limited external resource ($LPP$ model) are introduced by Gutiérrez et al. (2016, 2017). Gutiérrez et al. (2106) define $LPP$ games with externalities (Thrall and Lucas, 1963) but without considering any allocation rule. Gutiérrez et al. (2107) study the existence of Nash equilibria in the context of a competitive mechanism of allocation. Funaki and Yamato (1999) define a game with externalities to analyze the distribution of fish among fishermen, where the demands for it are additive and all agents have the same concave production function depending on labor as the only input. On the other hand, bankruptcy techniques have been widely used to deal with scarce resources in a huge array of economic problems such as mobile radio networks (Lucas-Estañ et al., 2012), k-hop mínimum cost spanning tree problems (Bergantiños et al., 2012), project management (Estévez-Fernández, 2012), allocation of nitrate emissions (Kampas and White, 2003) and allocation of the total carbon budget (Giménez-Gómez et al., 2016). However, in none of the papers mentioned in this paragraph has the problem of the resource allocation and the effect of cooperation been analyzed as a whole, i.e., that the definition of the cooperative game is determined by the allocation rule chosen together with the characteristics of the problem. In this way, a different game is obtained for each allocation rule used. In this sense, this paper is, as far as we know, novel and interesting since the analysis is performed considering the possible results of the allocation when a oncrete distribution rule is used as a consequence of the players' own strategies regarding the allocation process. This brings an innovation to the analysis of carbon dioxide emission permit allocations and their effects on the (strategic) behavior of companies under policies of abatement of GHG emissions.

## 3 Our proposal

In the framework of a production situation we propose a model in which there is a cap, a tax -price per unit- and transfers between the agents involved are allowed. We should emphasize that it can be seen as a combination of the use of tax and the cap-and-trade systems. This is because we take into account the cap and consider a fixed price (tax) per unit as managerial elements. However, we do not give our attention to how the trade among the agents takes place, but try to guarantee that the market can achieve a stable distribution of the revenues. This is modeled through cooperative games with transferable utility.

In this model the cap is allocated by the authority to the firms by means of emission permits that have a fixed price per unit. Therefore, the carbon dioxide emission permits can be seen as one more resource in the production system, that can be obtained individually or by groups of firms in order to increase their profits. In this paper we assume that groups of firms, called coalitions, can coordinate in order to make their demands on the quota of carbon dioxide. If the total demand of the sector is greater than the cap, the authority must establish a sharing method in order to allocate the cap among the applicants.

By introducing a tax (fixed price per unit) paid by the polluting firms two important objectives in the Paris Agreement can be achieved: the authorities in each country can obtain funds -because they have to invest in research, development and transfer of new technologies- and provide financial support to developing countries, whose need for financial and technological support for the implementation of these commitments is established in the agreement.

Moreover, we allow for the practice of utility transfers among the agents involved. This can be seen as a kind of private trading system.

The underlying idea of our model is the following: when the limit of greenhouse gas emissions is not sufficient to satisfy the demands of the firms, the authority can use a bankruptcy rule, $f$, to obtain a reasonable distribution of allowances among the different groups of firms, which will use their permits to produce and optimize their profits. Hence, given an $LPP$ situation we assume that the manager of the carbon dioxide emission permits announces which bankruptcy rule $f$ will be used to share these and name this an $f - LPP$ situation. Depending on the benchmark bankruptcy rule selected (proportional ($PROP$), constrained equal awards ($CEA$), constrained equal losses ($CEL$), Talmud ($TAL$), etc.) different games with externalities will be defined. We propose to use rules such as the proportional, the Talmud or the constrained equal awards to avoid the possibility of a firm receiving nothing, which prevents it from producing. We focus on the $CEA$ rule in our examples for this reason and because, as we will show, it is incentive compatible, which means that the authority will be able to know the real needs of gas emissions for firms. We study the core of these games because we want to check whether the profits generated by the allocation of the carbon dioxide emission quotas are coalitionally stable.

# 4 Preliminaries

## 4.1 Cooperative TU-games

In cooperative game theory it is assumed that the agents (players) can commit to behaving in a socially optimal way. With this approach it is also assumed that players can make binding agreements. The different groups of players are referred to as coalitions. The coalitions that can be formed among the set of players, denoted by $N$ and named as the grand coalition, can enforce certain allocations. The main issue in cooperative game theory is to decide how the benefits obtained by cooperation should be shared among the players.

A characteristic of cooperative games is that players know that they can probably achieve a larger total profit by pooling their resources than by acting individually. In cooperative games with transferable utility, TU-games, the earnings of a coalition can be expressed as one number. It can be seen as an amount of utility (money) and the implicit assumption is that it can be transferred among the players, i.e., there are side-payments. In our context, this means we allow players to trade their goods and permits among each other, in order to achieve a better result for all those involved with respect to their individual situation. The reader is referred to the text by González-Díaz et al. (2010) for a detailed study of TU-games. Within the class of TU-games we will distinguish between two subclasses: games in characteristic function form and games in partition function form, also called games with externalities.

Let $N$ be a non empty finite set of $n$ agents who agree to coordinate their actions. A cooperative game in characteristic function form is an ordered pair $(N, v)$, where $N$ is the set of players and $v : 2^N \to \mathbb{R}$ is the characteristic function with $v(\emptyset) = 0$. This function assigns to each group of players (coalition), $S \subset N$, the value $v(S)$ which represents what the members in $S$ obtain when they cooperate jointly.

In cooperative game theory we are interested in knowing how to share the joint profit among the cooperating agents. The core, $C(v)$, of a characteristic function form game $(N, v)$ is the set of distributions of $v(N)$ upon each coalition $S$ will receive at least as much it can obtain on its own, i.e., the subset of vectors in $\mathbb{R}^N$ satisfying

$$\begin{array}{ll} \text{(Efficiency)} & \sum_{i \in N} x_i = v(N), \text{ and} \\ \text{(Coalitional rationality)} & \sum_{i \in S} x_i \geq v(S), \text{ for all } S \subset N. \end{array}$$

Let $\mathcal{P}(N)$ denote the set of all partitions of $N$ and $P = \{S_1, \ldots, S_k\}$ represent one of these partitions, where the coalitions $S_1, \ldots, S_k$ are disjoint and their union is $N$. A cooperative game in partition function form is defined by $\left(N, \mathcal{P}(N), \{V(\bullet| P)\}_{P \in \mathcal{P}(N)}\right)$, where $N$ is the set of players (firms), $\mathcal{P}(N)$ denotes the set of all partitions of $N$ and $V(S| P)$ with $S \in P$ is a real number that represents the profit that a coalition $S \subset N$ can obtain when $P$ is formed. Note that the profit that a coalition can obtain depends on the coalitions formed by the other players in $P \in \mathcal{P}(N)$, therefore there are externalities.

Given a partition $P \in \mathcal{P}(N)$, a vector $x \in \mathbb{R}^n$ is said to be feasible under $P$ if it satisfies $\sum\limits_{i \in S} x_i \leq V(S|P), \forall S \in P$. We denote by $\mathcal{F}^P$ the set of all feasible vectors under $P$ and $\mathcal{F} = \cup_{P \in \mathcal{P}(N)} \mathcal{F}^P$ denotes the set of all feasible vectors. Given two vectors $x, x'$ in $\mathbb{R}^n$, we say that $x$ dominates $x'$ through $S$ and denote $x\ dom_S\ x'$ if the following conditions are satisfied:

1. $\sum\limits_{i \in S} x_i \leq V(S|P), \forall P \in \mathcal{P}(N)$ such that $S \in P$,

2. $x_i > x'_i, \forall i \in S$.

We say that $x$ dominates $x'$ if there exists $S \subset N$ such that $x\ dom_S\ x'$, and denote $x\ dom\ x'$. The core of a cooperative game in partition function form is defined by $C(V) = \{x \in \mathcal{F} \,|\nexists x' \in \mathcal{F} \text{ s.t. } x'\ dom\ x\}$. However, if we consider another definition of dominance, then we will obtain a different core. Thus, if we change condition 1 by

$$\overline{1}. \sum_{i \in S} x_i \leq V(S|P), \text{ for some } P \in \mathcal{P}(N) \text{ with } S \in P,$$

we obtain a more restrictive concept of dominance that we denote by $\overline{dom}$ and the corresponding core is defined as $\overline{C}(V) = \left\{x \in \mathcal{F} \left|\nexists x' \in \mathcal{F} \text{ s.t. } x'\ \overline{dom}\ x\right.\right\}$.

Associated with each game in partition function form two cooperative games in characteristic function form can arise: $(N, v^-)$ and $(N, v^+)$, where

$$v^-(S) = \min\{V(S|P) \,|P \in \mathcal{P}(N) \text{ such that } S \in P\},$$
$$v^+(S) = \max\{V(S|P) \,|P \in \mathcal{P}(N) \text{ such that } S \in P\}.$$

$(N, v^-)$ represents a pessimistic point of view regarding the gain that a coalition $S$ can attain, while $(N, v^+)$ can be seen as its optimistic counterpart. Funaki and Yamato (1999) proved that if $V(\{N\}|N) > \sum\limits_{S \in P} V(S|P), \forall P \in \mathcal{P}(N)$, then $C(V) = C(v^-)$ and $\overline{C}(V) = C(v^+)$.

## 4.2 Bankruptcy problems

A standard bankruptcy problem can be described by a triple $(N, E, d)$, where $N = \{1, ..., n\}$ is the finite set of agents, $E \geq 0$ is the estate to be divided and $d \in \mathbb{R}^N_+$, the vector of claims, is such that $\sum_{i \in N} d_i \geq E$. Every standard bankruptcy problem $(N, E, d)$ gives rise to a standard bankruptcy game $(N, v)$, where the value of a coalition $S \subset N$ is given by

$$v(S) = \max\{E - \sum_{i \in N \setminus S} d_i, 0\},$$

and represents what is left for players in $S$ after the demands of the players in $N \setminus S$ have been satisfied. These games have a non empty core.

A bankruptcy rule is a function $f$ that assigns to every bankruptcy problem $(N, E, d)$ a vector $f(N, E, d) \in \mathbb{R}^N$ such that $0 \leq f_i(N, E, d) \leq d_i$ for all $i \in N$, and $\sum_{i \in N} f_i(N, E, d) = E$.

In particular, we will focus on the $CEA$ rule, where $CEA_i(N, E, d) = \min\{d_i, \lambda\}$, for all $i \in N$, and $\lambda$ such that $\sum_{i \in N} CEA_i(N, E, d) = E$. We also use the proportional rule that divides the estate proportionally to the claims, i.e., $\frac{E}{\sum_{i \in N} d_i} d_j, j \in N$.

## 4.3 Production model description and assumptions

We consider production situations where several firms own resource bundles that can be used to turn out various products. Their production functions are such that the inputs must be combined in fixed proportions, the so-called Leontief production function which is one of those most often used in production literature. Furthermore, all firms have the same production function but differ in the amount of resources which they can manage. The goal of each firm is to maximize their profit, which equals the revenue from their products at the given market prices. We also consider that there is an external resource, carbon dioxide emission permits, limited by a certain amount, that agents need to obtain for producing their goods and must pay a tax for its consumption.

Formally, let $N = \{1, \ldots, n\}$ be a set of firms that address a production problem to produce a set $G = \{1, \ldots, g\}$ of goods from a set $Q = \{1, \ldots, q\}$ of resources. There is an external resource, limited by an amount of $r$, that agents need to obtain for producing the goods. A linear production situation with a limited external resource ($LPP$ situation) can be described by $(A, B, p, r, c)$, where

1. $A \in \mathcal{M}_{(q+1) \times g}$ is the production matrix, $a_{tj}$ represents the amount of the resource $t$ needed to produce item $j$, where the last row corresponds to the limited external resource and $a_{(q+1)j} > 0, \forall j \in G$, and there is at least one resource $t \in Q$ with $a_{tj} > 0, \forall j \in G$.

2. $B \in \mathcal{M}_{q \times n}$ is the resource matrix, where $b^i \in \mathbb{R}^q_+$ are the available resources for firm $i \in N$ and $b^S = \sum_{i \in S} b^i$. We assume that there is a positive quantity available of each resource, that is, for all resources $t \in Q$ there is a firm $i$ such that $b^i_t > 0$.

3. The limited resource, managed by an authority, has a cost per unit $c$ and the cap is denoted by $r$, with $c, r > 0$.

4. The price vector is $p \in \mathbb{R}^g_{++}$ and we assume that $p_j > a_{(q+1)j}c, \forall j \in G$, in order to deal with a profitable process.

We should point out that the cost per unit $c$ plays the role of a tax on polluting. Likewise it is an element with which to control pollution emissions, because a higher cost for polluting encourages firms to develop cleaner technologies for producing certain goods. Otherwise, the price of those goods should be very high (see Condition 4) and then probably non competitive. At the same time, since it is paid for by the pollutant companies, it also enables the authorities in each country to raise funds which can be invested in research, development and the transfer of new technologies and also provide financial support to developing countries.

If a coalition $S \subset N$ of firms cooperates, to maximize their profits they need an optimal production plan $(x; z) \in \mathbb{R}_+^{g+1}$ that provides information regarding how much of each product, $x$, they should produce and how many of the carbon dioxide emissions permits, $z$, they need. Not all production plans are feasible since the firms have to take into account their limited amount of resources. The amount of resources needed in a feasible production plan should not exceed the amount of resource available to cooperating firms. The following linear program maximizes the profit of coalition $S$

$$\begin{array}{ll} \max & \sum_{j=1}^{g} p_j x_j - cz \\ \text{s.t:} & Ax \leq \begin{pmatrix} b^S \\ z \end{pmatrix} \\ & x \geq \mathbf{0}_g, z \geq 0. \end{array} \tag{1}$$

The value of this linear program is denoted by $value\,(S; z)$, for every fixed amount of gas emission $z$.

The optimal demand of the carbon dioxide emission permits for each coalition $S$, $d_S = \min\{z \in \mathbb{R}_+ \,| value\,(S; z) \text{ is maximum}\}$, is obtained by solving the linear program (1). These optimal demands represent the desired amount of the carbon dioxide emission permits for each coalition $S$ and can be seen as their utopic or greatest aspirations a priori, i.e., before the carbon dioxide emission quota is allocated.

Although it may seem that the demands are superadditive, i. e., $d_S \geq \sum_{i \in S} d_{\{i\}}$, this is not true as Example 1 in Gutiérrez et al. (2016) shows.

Let us assume that $P$ is formed and the carbon dioxide emission permits finally allocated to coalition $S \in P$ by the manager is $z_S(P)$. The profit that a coalition $S \subset N$ can obtain is given by $value\,(S; z_S\,(P))$.

# 5 An embedded model of production situations and bankruptcy problems

In the previous section we have justified the role of a cost per unit for emission rights. But we also have a cap on the total emissions and therefore the rights of gas emission must be distributed. In the process of distributing the carbon dioxide emission permits, two cases can arise. If it is sufficient to satisfy the demands of the firms; everybody can be fully satisfied. However, we will consider the case where the firms claim more than the cap. In this scenario it is possible to enforce arbitration as in bankruptcy or rationing problems. In this section we will explain how controlling the cap can help in decreasing the GHG emissions considered.

Given a partition $P = \{S_1, \ldots, S_k\}$ of $N$, in this section we will consider that the elements in $P$ can claim more than the cap, $d(P) = \sum_{h=1}^{k} d_{S_h} > r$. Therefore, if this amount is exceeded the problem that faces partition $P$ can be modeled as a non standard bankruptcy problem, in the sense that $d_S \neq \sum_{i \in S} d_i$, in general. In this problem the authority should divide the amount $r$ (the estate) of the carbon dioxide emission permits among the set of coalitions according to the vector of demands, $(d_{S_1}, ..., d_{S_k})$, associated with the partition. Thus, it can be represented by $(P, r, (d_{S_1}, ..., d_{S_k}))$. Note that when we consider partition $P = \{\{i\}_{i \in N}\}$, although the bankruptcy problem is not a standard case it could be processed as a classic bankruptcy problem by considering that the claims are additive.

When $d(P) > r$ the authority can use bankruptcy rules to obtain a reasonable distribution of $r$ among the different coalitions, which will use their share to produce and optimize their profits. Let $(A, B, p, r, c)$ be an $LPP$ situation, we assume that the manager of the carbon dioxide emission permits announces which bankruptcy rule $f$ will use in order to allocate the quota of gas emissions. We call $(A, B, p, r, c, f)$ an $f - LPP$ situation. Depending on the rule selected (proportional ($PROP$), constrained equal awards ($CEA$), Talmud, recursive completion, etc.) different partition function form games are defined.

**Definition 1** *Let $(A, B, p, r, c, f)$ be an $f-LPP$ situation. The $f-LPP$ partition function form game associated with this situation is given by $\left(N, \mathcal{P}(N), \left\{V^f\left(\bullet | P\right)\right\}_{P \in \mathcal{P}(N)}\right)$, where $N$ is the set of players, $\mathcal{P}(N)$ denotes the set of all partitions of $N$ and $V^f\left(S | P\right)$ is obtained from*

$$\begin{array}{ll} \max & \sum_{j=1}^{g} p_j x_j - cf\left(S \left| P\right.\right) \\ s.t: & Ax \leq \left( \begin{array}{c} b^S \\ f\left(S \left| P\right.\right) \end{array} \right) \\ & x \geq \mathbf{0}_g, \end{array} \tag{2}$$

*for all $S \subset N$, using the quota of the carbon dioxide emissions that $S$ has obtained applying the bankruptcy rule $f\left(P, r, (d_S, d_{S_1}, ..., d_{S_k})\right)$ in partition $P \in \mathcal{P}(N)$ with $S \in P$, i.e., $f\left(S \left| P\right.\right)$.*

Note that $V^f\left(S | P\right) = value\left(S; f\left(S \left| P\right.\right)\right)$. When there is no possibility of confusion, we denote the $f - LPP$ game in partition function form by $\left(N, \mathcal{P}(N), V^f\right)$.

Similarly to Funaki and Yamato (1999) we need to prove, in our context, the next result to simplify the study of the core, i.e., in order to obtain coalitionally stable allocations.

**Proposition 2** *Let $(A, B, p, r, c, f)$ be an $f-LPP$ situation and $\left(N, \mathcal{P}(N), V^f\right)$ the corresponding partition function form game. Then,*

$$V^f\left(\{N\} | N\right) \geq \sum_{S \in P} V^f\left(S | P\right), \forall P \in \mathcal{P}(N).$$

As a result of the previous proposition, it can be derived that $C(V^f) = C(v_f^-)$ and $\overline{C}\left(V^f\right) = C(v_f^+)$, where

$$v_f^+(S) = \max_{P:S\in P} V^f\left(S|P\right) \text{ and } v_f^-(S) = \min_{P:S\in P} V^f\left(S|P\right) \tag{3}$$

are the optimistic and the pessimistic games associated with the $f-LPP$ game in partition function form, $\left(N, \mathcal{P}(N), V^f\right)$. We do not explicitly give the proof, because taking into account Proposition 2 it follows the same scheme of reasoning as in Funaki and Yamato (1999).

In order to illustrate how the externalities work for obtaining the game $\left(N, \mathcal{P}(N), V^f\right)$ the next example is given where the $CEA$ rule is used. Although it is possible to use any bankruptcy rule, the choice would depend on the characteristics of the problem and the properties of the rule that is used. We should point out that the data used in the examples throughout the paper are not from real-life contexts, but are only for illustrative purposes.

**Example 3** *Let $(A, B, r, p, c)$ be an LPP situation, with three firms, $N = \{1, 2, 3\}$, which produce two products from two resources and limited carbon dioxide emission permits, where*

$$A = \begin{bmatrix} 2 & 3 \\ 3 & 2 \\ 1 & 1 \end{bmatrix}, B = \begin{bmatrix} 40 & 60 & 80 \\ 60 & 40 & 50 \end{bmatrix}, p = \begin{pmatrix} 50 \\ 60 \end{pmatrix}, c = 14, r = 50$$

*The possible partitions in $N$ and their associated demands are*

$$P^1 = \{\{1\}, \{2\}, \{3\}\}, d^1 = (20, 20, 25), \quad P^2 = \{\{1, 2\}, \{3\}\}, d^2 = (40, 25),$$
$$P^3 = \{\{1, 3\}, \{2\}\}, d^3 = (46, 20), \quad P^4 = \{\{2, 3\}, \{1\}\}, d^4 = (45, 20),$$
$$P^5 = \{\{1, 2, 3\}\}, d^5 = 66.$$

*If we apply the $CEA$ rule to every bankruptcy problem associated with the previous partitions, we obtain*

$$CEA\left(P^1, 50, d^1\right) = (16.67, 16.67, 16.67), \quad CEA\left(P^2, 50, d^2\right) = (25, 25),$$
$$CEA\left(P^3, 50, d^3\right) = (30, 20), \quad CEA\left(P^4, 50, d^4\right) = (30, 20),$$
$$CEA\left(P^5, 50, d^5\right) = 50.$$

*Using these amounts in their own production processes, each coalition will attain*

$$V^{CEA}\left(\{1\}|\, P^1\right) = 666.67, \quad V^{CEA}\left(\{2\}|\, P^1\right) = 766.67, \quad V^{CEA}\left(\{3\}|\, P^1\right) = 766.67,$$
$$V^{CEA}\left(\{1, 2\}|\, P^2\right) = 1150, \quad V^{CEA}\left(\{3\}|\, P^2\right) = 1150,$$
$$V^{CEA}\left(\{1, 3\}|\, P^3\right) = 1380, \quad V^{CEA}\left(\{2\}|\, P^3\right) = 920,$$
$$V^{CEA}\left(\{2, 3\}|\, P^4\right) = 1380, \quad V^{CEA}\left(\{1\}|\, P^4\right) = 720,$$
$$V^{CEA}\left(\{1, 2, 3\}|\, P^5\right) = 2300.$$

*Note that this is a game with externalities since, for instance, what agent 1 receives in partition $P^1$, $V^{CEA}\left(\{1\}|\, P^1\right)$, is different from what obtains in partition $P^4$, $V^{CEA}\left(\{1\}|\, P^4\right)$, and this is because what agent 1 obtains depend on not only its own coalition, $\{1\}$, but also on how outsiders are organized: players 2 and 3 acting separately, $\{\{2\}, \{3\}\}$ or together, $\{2, 3\}$.*

Using the optimistic and pessimistic characteristic function games, introduced in (3), the optimistic and pessimistic cores are defined.

**Definition 4** *Let $(N, \mathcal{P}(N), V^f)$ be an $f-LPP$ game in partition function form. The optimistic core is defined by*

$$C\left(v_f^+\right) = \left\{x \in \mathbb{R}^n |\ x(S) \geq v_f^+(S)\ \forall S \text{ and } x(N) = V^f(N)\right\} = \overline{\mathcal{C}}\left(V^f\right).$$

*The pessimistic core is*

$$C\left(v_f^-\right) = \left\{x \in \mathbb{R}^n |\ x(S) \geq v_f^-(S)\ \forall S \text{ and } x(N) = V^f(N)\right\} = \mathcal{C}\left(V^f\right).$$

The optimistic core is included in the pessimistic one, thus if it is non empty the pessimistic core is also non empty. But, on many occasions, it is empty as the next example illustrates. This means that the optimistic core represents a higher level of demand for players with respect to what they think that should obtain.

**Example 5** *Let $(A, B, r, p, c, CEA)$ be the $CEA-LPP$ situation described in Example 3 and $(N, \mathcal{P}(N), V^{CEA})$ its related game. In this case, the optimistic core will be the set of $x \in \mathbb{R}^3$ such that*

$$\begin{aligned} &x(\{1\}) \geq 720, x(\{2\}) \geq 920, x(\{3\}) \geq 1150, \\ &x(\{1,2\}) \geq 1150, x(\{1,3\}) \geq 1380, x(\{2,3\}) \geq 1380, x(\{1,2,3\}) = 2300. \end{aligned}$$

*But this set is empty, since $720 + 920 + 1150 > 2300$, thus, $\overline{\mathcal{C}}\left(V^{CEA}\right) = \varnothing$. However, it is easy to check that $(700, 800, 800)$ belongs to the pessimistic core*

$$\mathcal{C}\left(V^{CEA}\right) = \left\{ x \in \mathbb{R}^3 \middle|\ \begin{array}{l} x(\{1\}) \geq 666.67, x(\{2\}) \geq 766.67, x(\{3\}) \geq 766.67, \\ x(\{1,2\}) \geq 1150, x(\{1,3\}) \geq 1380, x(\{2,3\}) \geq 1380, \\ x(\{1,2,3\}) = 2300 \end{array} \right\}.$$

Let $\left(N, v_f^+\right)$ be the optimistic $f-LPP$ game for all $f$. If $d_N \leq r$ it can be derived that $\overline{\mathcal{C}}\left(V^f\right) \neq \varnothing$, using similar arguments to those in Gutiérrez et al. (2016) and taking into account that $V^f(S|P)$ is obtained from (2). If $d_N > r$ the optimistic core can be empty as the previous example shows. In order to assure the nonemptiness of these cores, we need to add some conditions.

The optimistic and pessimistic games induce two resource allocation games $\left(N, R_f^+\right)$ and $\left(N, R_f^-\right)$ in the following way:

- $\arg \max\limits_{P: S \in P} V^f(S|P) = \mathcal{M}_S^+$.

Note that $\mathcal{M}_S^+$ is a set of partitions. Then, for every $Q \in \mathcal{M}_S^+$, $f(S|Q)$ is obtained. We take $Q$ such that $f(S|Q)$ is minimum and define $R_f^+(S) = f(S|Q)$. Among all partitions that provide the maximum optimal value we choose the most efficient, i.e., that which produces the smallest amount of gas emissions.

- $\arg\min_{P:S\in P} V^f(S|P) = \mathcal{M}_S^-$.

As we mention above $\mathcal{M}_S^-$ is a set of partitions. Then, for every $Q \in \mathcal{M}_S^-$, $f(S|Q)$ can be derived. We consider $Q$ such that $f(S|Q)$ is minimum and define $R_f^-(S) = f(S|Q)$.

We should highlight the key role that these resource allocation games play to establish in Theorem 7 that if the cores of the allocation games are non empty, then the cores of $\left(N, v_f^+\right)$ and $\left(N, v_f^-\right)$ are non empty when $d_N > r$. They are focused on the cap and, therefore, on what the agents (firms) will demand from the authority and receive according to the bankruptcy rule used. Thus, they have a great impact on profits. Clearly, $\left(N, R_f^+\right)$ and $\left(N, R_f^-\right)$ depend on the bankruptcy rule $f$, the partitions and the linear production with a global cap on carbon dioxide gas emissions. Likewise, $R_f^+(S) \geq R_f^-(S), \forall S \subset N$. This implies that $C\left(R_f^+\right) \subset C\left(R_f^-\right)$ and $C\left(R_f^+\right)$ is more demanding than $C\left(R_f^-\right)$. The next example illustrates this fact, where $C\left(R_f^-\right) \neq \varnothing$ and $C\left(R_f^+\right) = \varnothing$.

**Example 6** *Let $(A, B, r, p, c, CEA)$ be the $CEA - LPP$ situation described in Example 3. The corresponding $\left(N, R_{CEA}^+\right)$ is given by*

$$R_{CEA}^+(\{1\}) = 20, R_{CEA}^+(\{2\}) = 20, R_{CEA}^+(\{3\}) = 25$$
$$R_{CEA}^+(\{1,2\}) = 25, R_{CEA}^+(\{1,3\}) = 30, R_{CEA}^+(\{2,3\}) = 30, R_{CEA}^+(N) = 50$$

*and $\left(N, R_{CEA}^-\right)$ is*

$$R_{CEA}^-(\{1\}) = 16.67, R_{CEA}^-(\{2\}) = 16.67, R_{CEA}^-(\{3\}) = 16.67$$
$$R_{CEA}^-(\{1,2\}) = 25, R_{CEA}^-(\{1,3\}) = 30, R_{CEA}^-(\{2,3\}) = 30, R_{CEA}^-(N) = 50.$$

*Note that the core of $\left(N, R_{CEA}^+\right)$ is empty, while the core of $\left(N, R_{CEA}^-\right)$ is not empty since $(16.67, 16.67, 16.67) \in C\left(R_{CEA}^-\right)$.*

The next result is relevant because it states that if the manager of the gas emission permits establishes a stable allocation of these among the players, then a stable distribution of the revenues can be obtained. This can be done by means of a permits market and no agent can complain about the allocations of permits or the profits achieved.

**Theorem 7** *Let $(A, B, p, r, c, f)$ be an $f - LPP$ situation and $\left(N, \mathcal{P}(N), V^f\right)$ the corresponding partition function form game such that $d_N > r$. Then $C\left(R_f^-\right) \neq \varnothing$ (resp. $C\left(R_f^+\right) \neq \varnothing$) implies $C\left(V^f\right) \neq \varnothing$ (resp. $\overline{C}\left(V^f\right) \neq \varnothing$).*

In the next corollary we require $\sum_{i\in N} d_i > r$ to have a bankruptcy problem $\left(N, r, (d_i)_{i\in N}\right)$, in order to assure the nonemptiness of the core. This condition plays a key role and, in our benchmark, it can be fulfilled by reducing the cap.

**Corollary 8** *Let $(A, B, p, r, c, f)$ be an $f - LPP$ situation and $\left(N, \mathcal{P}(N), V^f\right)$ the corresponding partition function form game, such that $d_N > r$ and $\sum_{i \in N} d_i > r$. If $f\left(N, r, (d_i)_{i \in N}\right) \in C\left(R_f^-\right)$ (resp. $f\left(N, r, (d_i)_{i \in N}\right) \in C\left(R_f^+\right)$), then $C\left(V^f\right) \neq \varnothing$ (resp. $\overline{C}\left(V^f\right) \neq \varnothing$).*

We should stress the importance of this result which enables us to solve a difficult problem obtaining a stable allocation in a straightforward way. Because on setting the cap the authority can decrease this until it falls below the sum of the individual demands of the firms. Then, all firms will cooperate and if the authority uses an allocation in the core of the resource allocation game, then we are able to find a core element à la Owen using duality and a bankruptcy rule. This guarantees at least one stable distribution of the revenue but in general, there are more.

Condition $f\left(N, r, (d_i)_{i \in N}\right) \in C\left(R_f^-\right)$ cannot be neglected as the next example shows, where, in particular, the proportional rule is used.

**Example 9** *Let $(A, B, r, p, c)$ be the LPP situation described in Example 3. When we apply the proportional $(PROP)$ rule to every bankruptcy problem associated with each partition, we obtain*

$$\begin{array}{ll} PROP\left(P^1, 50, d^1\right) = (15.38, 15.38, 19.23), & PROP\left(P^2, 50, d^2\right) = (30.77, 19.23), \\ PROP\left(P^3, 50, d^3\right) = (34.85, 15.15), & PROP\left(P^4, 50, d^4\right) = (34.62, 15.38), \\ PROP\left(P^5, 50, d^5\right) = 50. & \end{array}$$

*Using these amounts in their own production processes, the worth of each coalition is given by*

$$\begin{array}{lll} V^{PROP}\left(\{1\} | P^1\right) = 646.08, & V^{PROP}\left(\{2\} | P^1\right) = 707.48, & V^{PROP}\left(\{3\} | P^1\right) = 884.58, \\ V^{PROP}\left(\{1,2\} | P^2\right) = 1415.42, & V^{PROP}\left(\{3\} | P^2\right) = 884.58, & \\ V^{PROP}\left(\{1,3\} | P^3\right) = 1603.10, & V^{PROP}\left(\{2\} | P^3\right) = 696.90, & \\ V^{PROP}\left(\{2,3\} | P^4\right) = 1592.52, & V^{PROP}\left(\{1\} | P^4\right) = 646.08, & \\ V^{PROP}\left(\{1,2,3\} | P^5\right) = 2300. & & \end{array}$$

*In this case, $d_N > 50$ and $\sum_{i \in N} d_i > 50$. However,*

$$\begin{array}{l} R_{PROP}^-(\{1\}) = 15.38, R_{PROP}^-(\{2\}) = 15.15, R_{PROP}^-(\{3\}) = 19.23 \\ R_{PROP}^-(\{1,2\}) = 30.77, R_{PROP}^-(\{1,3\}) = 34.85, R_{PROP}^-(\{2,3\}) = 34.62, R_{PROP}^-(N) = 50. \end{array}$$

*and $(15.38, 15.38, 19.23)$ does not belong to $C\left(R_{PROP}^-\right)$ because it is empty. In this case,*

$$\begin{array}{lll} v_{PROP}^-(\{1\}) = 646.08, & v_{PROP}^-(\{2\}) = 696.90, & v_{PROP}^-(\{3\}) = 884.58, \\ v_{PROP}^-(\{1,2\}) = 1415.42, & v_{PROP}^-(\{1,3\}) = 1603.10, & v_{PROP}^-(\{2,3\}) = 1592.52, \\ v_{PROP}^-(\{1,2,3\}) = 2300 & & \end{array}$$

*and $C\left(v_{PROP}^-\right) = \varnothing$.*

The implications of instability in the allocation of permits can lead to instability in the distribution of revenues as the previous example shows. Thus, the system will produce complaints from the agents regarding the allocation of carbon dioxide emissions. As a result, dissatisfaction with the allocation of permits will arise. This leads us to seek rules in which the allocation of permits is stable and therefore provides a stable distribution of profits.

# 6 Allocation of the carbon dioxide emission permits using the CEA rule

In this section we focus on the $CEA$ rule for several reasons. First of all, because it benefits firms with lower gas emissions needs; while for instance, the constrained equal losses rule benefits those with higher demands. Secondly, because as we will prove, under certain conditions, $CEA\left(N, r, (d_i)_{i\in N}\right) \in C\left(R_f^-\right)$ which assures a stable allocation of profits. Thirdly, we should mention that it has the merging proofness property[1], i.e., if $k, j \in N$ join, then

$$CEA_{kj}\left(N^*, r, (d_i^*)_{i\in N^*}\right) \leq CEA_k\left(N, r, (d_i)_{i\in N}\right) + CEA_j\left(N, r, (d_i)_{i\in N}\right),$$

where $N^* = N \backslash \{k, j\} \cup \{kj\}$. This means that when two firms join they are not better off, which implies the non manipulability by unions, i.e., this rule is immune to strategic manipulations when a group of firms merge in order to be represented as a single firm. The reader is referred to Thomson (2015) for more details. Finally, as we will prove in subsection 6.2 , as a direct mechanism it is incentive compatible, truth-telling or strategy-proofness.

## 6.1 Stability

The pessimistic core is non empty when we consider the $CEA$ rule, under certain assumptions, as Corollay 13 states. In order to prove this we need a previous result.

**Lemma 10** *Let $(A, B, r, p, c, f)$ be a $f-LPP$ situation with $d_N > r$ and $\sum_{i\in N} d_i \geq r$. Given $S \subset N$, if $R_f^-(S) = d_S$ then*

$$\arg\min_{P:S\in P} V^f(S\,|P) = \{P \in \mathcal{P}(N)\,|S \in P\}.$$

This result guarantees that if a coalition in the pessimistic case obtains all the needs, then all partitions give the same revenue.

The next theorem means that if there is no possibility of manipulation by merging, the core of the pessimistic allocation game is non empty.

[1] Although the proportional rule is the only non manipulable bankruptcy rule in standard bankruptcy problems, it fails to be non manipulable in this context. In Example 9 $R_{PROP}^-(\{1\}) + R_{PROP}^-(\{3\}) = 15.38 + 19.23 = 34.71 < 34.85 = R_{PROP}^-(\{1,3\})$.

**Theorem 11** *Let* $(A, B, r, p, c, f)$ *be a* $f - LPP$ *situation with* $d_N > r$ *and* $\sum_{i \in N} d_i \geq r$. *If* $\sum_{i \in S} f_i \left(N, r, (d_i)_{i \in N}\right) \geq f_S \left(S \cup \{i\}_{i \in N \setminus S}, r, \left(d_S, (d_i)_{i \in N \setminus S}\right)\right) = f\left(S \middle| S \cup \{i\}_{i \in N \setminus S}\right), \forall S \subset N$, *then* $C\left(R_f^-\right) \neq \varnothing$.

The following result establishes a somewhat of weak merging proofness property for the $CEA$ rule in our framework.

**Proposition 12** *Let* $(A, B, r, p, c, CEA)$ *be a* $CEA - LPP$ *situation with* $d_N > r$ *and* $d_i + d_j \geq \frac{2r}{n}$, *for all* $i, j \in N$. *Then* $\forall S \subset N$,

$$CEA(S) = \sum_{i \in S} CEA\left(N, r, (d_i)_{i \in N}\right) \geq$$
$$CEA_S\left(\left\{S \cup \{i\}_{i \in N \setminus S}, r, \left(\sum_{i \in S} d_i, (d_i)_{i \in N \setminus S}\right)\right\}\right) = CEA\left(S \middle| S \cup \{i\}_{i \in N \setminus S}\right).$$

**Corollary 13** *Let* $(A, B, r, p, c, CEA)$ *be a* $CEA - LPP$ *situation with* $d_N > r$ *and* $d_i + d_j \geq \frac{2r}{n}$, *for all* $i, j \in N$. *Then* $C\left(V^f\right) \neq \varnothing$.

This result can help the authority to manage the cap on greenhouse gas emissions in order to deal with progressive reductions. Since making it less than the sum of the individual demands and applying the $CEA$ rule a stable allocation can be derived without difficulty by using duality (see Corollary 8 and Corollary 13). Although this can be an extreme allocation as can be seen in the next example. In consequence, no complaints about the allocation of permits will arise and therefore, we will attain an environmentally sustainable system.

**Example 14** *Let* $(A, B, r, p, c, CEA)$ *be the* $CEA - LPP$ *situation described in Example 3 and* $\left(N, \mathcal{P}(N), V^{CEA}\right)$ *its related game* $CEA - LPP$ *game. We have seen that* $(16.67, 16.67, 16.67) \in C\left(R_{CEA}^-\right)$. *An optimal solution for the dual program of the grand coalition linear problem is* $y_1^* = y_2^* = 0$ *and* $y_3^* = 60$. *Thus, the allocation* $(766.67, 766.67, 766.67) \in C\left(v_{CEA}^-\right)$. *But this distribution of the profits is extreme, because all extra revenue derived from cooperation goes to player 1.*
*However, the core is bigger, for instance,* $(700, 800, 800) \in C\left(v_{CEA}^-\right)$. *Assuming that the profits are in euros, this result can be obtained from the initial distribution of the resource* $(16.67, 16.67, 16.67)$ *by means of trading the permits among agents in the following way:*
*Player 1 sells* $3\frac{1}{3}$ *units to players 2 and 3 at* 50 *euros per unit. Thus, player 1 will obtain* 600 *euros from her own production using 10 units. She has paid* $16\frac{2}{3} \times 14$ *for buying the emission permits and received* $6\frac{2}{3} \times 50$ *for selling part of the permits. So, this will give her a profit of* 700 *euros.*
*Player 2 (respectively 3) will attain* 1200 *euros from her own production using 20 permits. She has to pay* $16\frac{2}{3} \times 14$ *to the manager for buying the emission permits and* $3\frac{1}{3} \times 50$ *for buying the permits to player 1. This will give her a net profit of* 800 *euros.*
*This example describes how from a stable allocation of the permits we can reach, through a market with side-payments, a stable distribution of the profits. This*

*has generated* 700 *euros for the manager, via taxes and controlling the emissions with a cap of* 50*, i.e., he has applied a quantitative control over gas emissions and has had no influence at all in the market. However, he has set up the basis for obtaining a stable result, as Corollaries 8 and 13 indicated.*

## 6.2 Incentive compatibility

A priori the manager does not know the real emissions, for this reason in this subsection we will focus on designing a mechanism that is fair for firms and obliges them to be truthful. In this way the manager will have information to know how much and how to reduce the cap to achieve the goal of sustainability and at the same time, to be fairer because it will based on real data which cannot be manipulated by the firms.

Let $\Theta$ be the set of all possible needs of carbon dioxide emissions. This is the set of types of firms. Let $N$ be the set of all firms requesting emission permits. Let $\Gamma$ be the set of all feasible allocations of carbon dioxide emissions.

An allocation rule (direct mechanism) is a function $A : \Theta^N \to \Gamma$.

Given an allocation rule $A$ and a vector of needs $\theta = (\theta_1, ..., \theta_1)$, the payment of firm $i$ is given by

$$\pi_i\left(A\left(\theta\right)\right) = value\left(i; A_i\left(\theta\right)\right).$$

Therefore, firms in $N$ are facing a non cooperative game in which all companies have the same set of alternatives, $\Theta$, each company knows its real necessity for carbon dioxide emission but does not know the needs of the other companies. In other words, that information is private. A strategy for company $i$ is a function from $\Theta$ into itself, $\sigma_i : \Theta \to \Theta$. The set of all strategies is given by $\Sigma$ (for all players).

**Definition 15** *A profile $\sigma^*$ is an equilibrium if*

$$\pi_i\left(A\left(\sigma^*\left(\theta\right)\right)|\theta_i\right) \geq \pi_i\left(A\left(\sigma_i\left(\theta_i\right); \sigma^*_{-i}\left(\theta_{-i}\right)\right)|\theta_i\right), \forall i \in N \text{ and } \forall \sigma_i \in \Sigma,$$

*where* $\sigma^*\left(\theta\right) = \left(\sigma^*_1\left(\theta_1\right), ..., \sigma^*_{i-1}\left(\theta_{i-1}\right), \sigma^*_i\left(\theta_i\right), \sigma^*_{i+1}\left(\theta_{i+1}\right), ..., \sigma^*_n\left(\theta_n\right)\right)$ *and* $\sigma^*_{-i}\left(\theta_{-i}\right) = \left(\sigma^*_1\left(\theta_1\right), ..., \sigma^*_{i-1}\left(\theta_{i-1}\right), \sigma^*_i\left(\theta_i\right), \sigma^*_{i+1}\left(\theta_{i+1}\right), ..., \sigma^*_n\left(\theta_n\right)\right)$.

**Definition 16** *A profile $\sigma^*$ is a dominant strategy equilibrium if the following holds*

$$\pi_i\left(A\left(\sigma^*_i\left(\theta_i\right); \sigma_{-i}\left(\theta_{-i}\right)\right)|\theta_i\right) \geq \pi_i\left(A\left(\sigma_i\left(\theta_i\right); \sigma_{-i}\left(\theta_{-i}\right)\right)|\theta_i\right), \forall i \in N, \forall \sigma_{-i} \in \Sigma_{-i} \text{ and } \forall \sigma_i \in \Sigma.$$

**Definition 17** *An allocation rule (direct mechanism), $A$, is incentive compatible (truth-telling or strategy-proofness) if the strategy profile* $\sigma^*\left(\theta\right) = \left(\sigma^*_1\left(\theta_1\right), ..., \sigma^*_n\left(\theta_n\right)\right) = \left(\theta_1, ..., \theta_1\right) = \theta$ *is a dominant strategy equilibrium of the game.*

**Theorem 18** *The CEA rule as a (direct) mechanism is incentive compatible.*

Thereby, applying this mechanism, firms have to be truthful when declaring their real needs for emission permits. The proportional rule, as a direct mechanism, does not fulfill the compatibility of incentives, because if an agent artificially increases her demand she will receive more emission permits. Therefore, there are incentives not to declare her real requirements.

# 7 Managerial implications

Our model provides the administration which issues the permits for the emission of greenhouse effect gases with two management tools: the cap and the tax or unit price of contaminant emission. The trade between companies can still take place but with the guarantee that stable results can be achieved. (See Example 14).

On the one hand, the cap allows the emissions to be reduced progressively, as should be done according to the Paris Agreement and on the other hand, fixing a price for the emissions permits guarantees two concerns. Firstly, that the most pollutant technologies will open the way for new less pollutant technologies, due to the increase in the production cost from the necessity to purchase a larger quantity of emission permits, which should encourage the introduction of these new technologies to replace the former. Secondly, this will provide income for governments to be able to allocate funds for two kinds of fundamental projects according to the Paris Agreement. In the first place, to finance R+D+I projects to improve technologies which can include the replacement of more pollutant technologies with others less pollutant as mentioned above, secondly, to finance projects which favour economic and environmental development in developing countries. Furthermore, the combination of a cap and a tax can reduce the possibility that abatement of emissions does not occur, because a suitable combination of both can be used.

For the management of the cap, the use of the bankruptcy rules is proposed, such as the $CEA$, the $TAL$ and the $PROP$, as they ensure that no-one is excluded in the production process since all the producers receive a positive quota of emission permits. Nevertheless, the $CEA$ rule has two advantages. The first is that it provides stable allocations of the emission permits which in turn, allows for stable distribution of the profit. The second is that it is strategy proofness, so that if the agency does not have the means to assess or is unaware of the companies' real pollutant gas emissions, the use of the $CEA$ rule ensures that companies will tell the truth regarding their demands for emission rights as there is nothing to be gained by cheating. Knowledge of the real emissions has another implication because, according to the Paris Agreement, the cap is to be revised every five years. Therefore, if the administration knows the agents' real demands, an emission level can be fixed, that is the cap, thereby making the allocation of the permits stable in the sense that no company or group of companies will feel that they have lost out, and with the trade between companies (let us remember that the transfer of the utility is permitted) a stable distribution of the profits generated can be reached. So this will also avoid complaints about the distribution of permits.

# 8 Concluding remarks

The model presented in this paper is an arbitration system, which takes into account a tax (fixed price per unit) and a limit (or cap) for the emission permits.

The tax is introduced bearing in mind that polluting technologies should be reformed. In addition, it also allows for obtaining funds to meet two purposes of the Paris Agreement: investment in research, development and transfer of new technologies and financial support for developing countries. The introduction of the cap is due to the fact that in the review process of the agreement the reduction of permits should be carried out in a progressive decreasing manner. This can be managed by the authority in such a way that the conditions of Corollary 8 are met. The utility transfers among firms is modeled by means of TU-games with externalities. Our model includes the essential aspects of the Paris Agreement and provides insights and tools in order to obtain a sustainable allocation of the permits that allows a fair distribution of the profits. Since with our bankruptcy approach we have been able to find a core element à la Owen in which the bankruptcy rule $f$, once it is fixed in Corollary 8, or using the $CEA$ rule in Corollary 13, takes part in the construction of the imputation. In addition, the $CEA$ rule as a mechanism meets the compatibility of incentives and therefore, companies have an incentive to declare their real needs of carbon dioxide.

In this paper we have considered all firms to have the same technology. We will study the case of different technologies such as those in linear transformation of product situations (Timmer et al., 2000) in further research.

The use of the $CEA$ rule in this framework was initially motivated by the fact that it always assigns a positive amount to each claimant, as is the case of the Talmud and the proportional rules. However, we have shown that the latter can give rise to dissatisfaction in the allocation of permits. Additionally, we have proved that the $CEA$ rule, as a direct mechanism, is incentive compatible in this context, but the proportional rule is not. Other rules could be used, but with a benchmark. For example, the $CEL$ rule could exclude those companies needing only a smaller share of carbon dioxide permits. Furthermore, as a direct mechanism it would not be incentive compatible. The $TAL$ rule, which is a combination of the $CEA$ rule and $CEL$ rule, has not been taken into account because neither of both parts of the rule is incentive compatible. The $CEL$ part of the $TAL$ rule is not incentive compatible because neither is the $CEL$ rule. Likewise, in the Paris Agreement progressive reductions on the cap should be made. Thus, in the end only the $CEA$ part of the $TAL$ rule would be implemented which is not incentive compatible, because it uses $\frac{d}{2}$ as the demands vector and this leads firms to be untruthful.

An extension of the standard bankruptcy model has arisen where claims are not additive and should be studied in further research. As in this case, other extensions of the standard model arising from real-life problems were introduced in Pulido et al. (2002, 2008), Casas-Méndez et al. (2011), Carpente et al. (2013) and Timoner and Izquierdo (2016).

In addition to the current approach this problem can be addressed through the design of two different models: with a non-cooperative point of view, as in Gutiérrez et al. (2017), or by using auction mechanisms, which will be studied in future research.

**Acknowledgements.** The authors thank two anonymous referees for their helpful comments and suggestions to improve the contents of the paper. Financial support from the Government of Spain (MICINN, MINECO) and FEDER under projects MTM2011-23205, MTM2011-27731-C03, MTM2014-53395-C3-3-P and MTM2014-54199-P and from Fundación Séneca de la Región de Murcia through grant 19320/PI/14 are gratefully acknowledged.

# 9 Appendix

**Proof of Proposition 2.** Given $P \in \mathcal{P}(N), V^f\left(S|\,P\right) = value\left(S; f\left(S\,|P\right)\right), \forall S \in P$. Let be $\left(x^S; f\left(S\,|P\right)\right)$ an optimal production plan for each coalition $S \in P$. Thus, $Ax^S \leq \begin{pmatrix} b^S \\ f\left(S\,|P\right) \end{pmatrix}$ and

$$A\left(\sum_{S\in P} x^S\right) \leq \begin{pmatrix} \sum\limits_{S\in P} b^S \\ \sum\limits_{S\in P} f\left(S\,|P\right) \end{pmatrix} \leq \begin{pmatrix} b^N \\ r \end{pmatrix}.$$

We distinguish two cases:

1) If $d_N \geq r, f\left(N\,|\{N\}\right) = r$, then $\left(\sum\limits_{S\in P} x^S; \sum\limits_{S\in P} f\left(S\,|P\right)\right)$ is a feasible production plan for $N$ and

$$\sum_{S\in P} value\left(S; f\left(S\,|P\right)\right) \leq value\left(N; \sum_{S\in P} f\left(S\,|P\right)\right) \leq V^f\left(N|\,\{N\}\right).$$

2) If $d_N < r$, then $f\left(N\,|\{N\}\right) = d_N$. Now $\left(\sum\limits_{S\in P} x^S; \sum\limits_{S\in P} f\left(S\,|P\right)\right)$ is a feasible solution of problem (1) for $N$. By definition of $d_N$, we have that

$$value\left(N; d_N\right) \geq \sum_{j=1}^{g} p_j\left(\sum_{S\in P} x^S\right) - c\sum_{S\in P} f\left(S\,|P\right) = \sum_{S\in P} value\left(S; f\left(S\,|P\right)\right).$$

■

**Proof of Theorem 7.** Let $y^*$ be an optimal solution of the dual problem of the grand coalition linear program

$$\begin{array}{ll} \max & \sum_{j=1}^{g} p_j x_j - cr \\ \text{s.t:} & Ax \leq \begin{pmatrix} b^N \\ r \end{pmatrix} \\ & x \geq \mathbf{0}_g. \end{array} \qquad (4)$$

It is easy to check that since $d_N > r$, then $y^*_{q+1} > c$. Let $h \in C\left(R^-_f\right)$, thus $\sum_{t=1}^{q} b^N_t y^*_t + ry^*_{q+1} - cr = v^-_f(N)$. Moreover, $\forall S \subset N$

$$\sum_{t=1}^{q} b^S_t y^*_t + \left(\sum_{i\in S} h_i\right) y^*_{q+1} - c\left(\sum_{i\in S} h_i\right) \geq$$
$$\sum_{t=1}^{q} b^S_t y^*_t + R^-_f(S)\left(y^*_{q+1} - c\right) \geq v^-_f(S),$$

where the last inequality holds because $y^*$ is feasible for the dual problem of coalition $S$ and $y^*_{q+1} > c$. Therefore,

$$\left( \sum_{t=1}^{q} b_t^i y_t^* + h_i \left( y^*_{q+1} - c \right) \right)_{i \in N} \in \mathcal{C}\left( V^f \right).$$

The same scheme can be used to derive the nonemptiness of $\overline{\mathcal{C}}\left( V^f \right)$. ■

**Proof of Corollary 8.** The proof follows the same lines as the proof of Theorem 7 if $f\left(N, r, (d_i)_{i\in N}\right) \in C\left(R_f^-\right)$ is used instead of $h \in C\left(R_f^-\right)$. ■

**Proof of Lemma 10.** If $Q \notin \arg\min_{P:S\in P} V^f(S|P)$, then $V^f(S|Q) > \min_{P:S\in P} V^f(S|P)$. But since $R_f^-(S) = d_S = \min\{z \in \mathbb{R}_+ \,|value(S;z) \text{ is maximum}\}$, $\min_{P:S\in P} V^f(S|P) = \max_{P:S\in P} V^f(S|P)$ and $V^f(S|Q) > \max_{P:S\in P} V^f(S|P)$, what is a contradiction. ■

**Proof of Theorem 11.** We will demonstrate that $f\left(N, r, (d_i)_{i\in N}\right) \in C\left(R_f^-\right)$.

Let $S \subset N$. We distinguish two cases:

*a)* $\left\{S \cup \{i\}_{i\in N\setminus S}\right\} \in \arg\min_{P:S\in P} V^f(S|P)$. This implies that $f\left(S \left| S \cup \{i\}_{i\in N\setminus S}\right.\right) \geq R_f^-(S)$, by definition of $R_f^-(S)$. Furthermore, by hypothesis $\sum_{i\in S} f_i\left(N, r, (d_i)_{i\in N}\right) \geq f_S\left(S \cup \{i\}_{i\in N\setminus S}, r, \left(d_S, (d_i)_{i\in N\setminus S}\right)\right) = f\left(S \left| S \cup \{i\}_{i\in N\setminus S}\right.\right) \geq R_f^-(S)$.

*b)* $\left\{S \cup \{i\}_{i\in N\setminus S}\right\} \notin \arg\min_{P:S\in P} V^f(S|P)$. We consider two situations:

*b*.1) If $d_S + \sum_{i\in N\setminus S} d_i \leq r$, then we have that

$$\sum_{i\in S} f_i\left(N, r, (d_i)_{i\in N}\right) \geq f_S\left(S \cup \{i\}_{i\in N\setminus S}, r, \left(d_S, (d_i)_{i\in N\setminus S}\right)\right) = d_S \geq R_f^-(S),$$

where the first inequality is obtained by hypothesis and the last is always fulfilled.

*b*.2) If $d_S + \sum_{i\in N\setminus S} d_i > r$, since $\left\{S \cup \{i\}_{i\in N\setminus S}\right\} \notin \arg\min_{P:S\in P} V^f(S|P)$, we have that $V^f\left(S \left| S \cup \{i\}_{i\in N\setminus S}\right.\right) > \min_{P:S\in P} V^f(S|P)$.

Let us assume that $f\left(S \left| S \cup \{i\}_{i\in N\setminus S}\right.\right) < R_f^-(S) < d_S$, where the last inequality holds by Lemma 10. This implies that $\exists \alpha \in (0,1)$ such that

$$R_f^-(S) = \alpha f\left(S \left| S \cup \{i\}_{i\in N\setminus S}\right.\right) + (1-\alpha)\, d_S.$$

Consider the solutions $\left(x^1; z^1\right)$ and $\left(x^2; z^2\right)$ for problem (2) with $f\left(S \left| S \cup \{i\}_{i\in N\setminus S}\right.\right)$ and $d_S$, respectively. Thus, $\alpha\left(x^1; z^1\right) + (1-\alpha)\left(x^2; z^2\right)$ is a feasible solution for

(2) with $R_f^-(S)$. Hence, we have that

$$\begin{aligned}
&\min_{P:S\in P} V^f(S|P) \geq \\
&\alpha V^f\left(S\left|S\cup\{i\}_{i\in N\setminus S}\right.\right) + (1-\alpha)\, value\,(S;d_S) > \\
&\min_{P:S\in P} V^f(S|P),
\end{aligned}$$

where the first inequality is due to the linearity of problem (2) and the last holds since $\left\{S\cup\{i\}_{i\in N\setminus S}\right\} \notin \arg\min_{P:S\in P} V^f(S|P)$ and the definition of $d_S$. Therefore we obtain a contradiction and

$$\sum_{i\in S} f_i\left(N,r,(d_i)_{i\in N}\right) \geq f\left(S\left|S\cup\{i\}_{i\in N\setminus S}\right.\right) \geq R_f^-(S).$$

Consequently, $f\left(N,r,(d_i)_{i\in N}\right) \in R_f^-(S)$. ■

**Proof of Proposition 12.** It is easy to check that $d_i + d_j \geq \frac{2r}{n}$, for all $i,j\in N$, implies $\sum_{i\in N} d_i \geq r$. Two cases can arise:

1) $d_S \geq \sum_{i\in S} d_i$. In this case, due to the merging proofness property

$$CEA(S) = \sum_{i\in S} CEA\left(N,r,(d_i)_{i\in N}\right) \geq CEA_S\left(\left\{S\cup\{i\}_{i\in N\setminus S}, r, \left(\sum_{i\in S} d_i, (d_i)_{i\in N\setminus S}\right)\right\}\right).$$

By definition of the $CEA$ rule, $CEA(S) \leq \sum_{i\in S} d_i$. Therefore,

$$CEA_S\left(\left\{S\cup\{i\}_{i\in N\setminus S}, r, \left(\sum_{i\in S} d_i, (d_i)_{i\in N\setminus S}\right)\right\}\right) \leq \sum_{i\in S} d_i$$

and by definition of the $CEA$ rule and the hypothesis, we have that

$$CEA_S\left(S\cup\{i\}_{i\in N\setminus S}, r, \left(\sum_{i\in S} d_i, (d_i)_{i\in N\setminus S}\right)\right) = CEA_S\left(S\cup\{i\}_{i\in N\setminus S}, r, \left(d_S, (d_i)_{i\in N\setminus S}\right)\right).$$

2) $d_S < \sum_{i\in S} d_i$.

We can distinguish two subcases:

2.1) Since $CEA(S) \geq d_S$, then the result directly holds.

2.2) $CEA(S) < d_S$, then we have that $\sum_{i\in N\setminus S} d_i + d_S > r$ since $\sum_{i\in N} d_i > r$. Furthermore, by the merging proofness property

$$\begin{aligned}
CEA(S) &\geq CEA_S\left(S\cup\{i\}_{i\in N\setminus S}, r, \left(\sum_{i\in S} d_i, (d_i)_{i\in N\setminus S}\right)\right) \\
&= CEA_S\left(S\cup\{i\}_{i\in N\setminus S}, r, \left(d_S, (d_i)_{i\in N\setminus S}\right)\right),
\end{aligned}$$

where the last equality holds because of the definition of the $CEA$ rule and $CEA(S) < d_S < \sum_{i\in S} d_i$. Thus, the result always holds. ■

**Proof of Corollary 13.** To prove the result we only have to take into account that using Theorem 11 and Proposition 12 together with Corollary 8, we can construct an element of the pessimistic core. ■

**Proof of Theorem 18.** Let $\{S_1, \ldots, S_k\}$ be a partition of the companies requesting carbon dioxide emission permits and let $d = (d_1, \ldots, d_k)$ be the vector of their real needs. Let $\Theta = [0, +\infty)$.

We consider the following strategy profile $(\sigma_1(d_1), ..., d_i, ..., \sigma_k(d_k))$ and distinguish two situations:

$a.1$. If $CEA_i(\sigma_1(d_1), ..., d_i, ..., \sigma_k(d_k)) < d_i$, then we have that $\sum_{j=1}^{k} \sigma_j(d_j) > r$.

Now, we distinguish three cases:

$a.1.1$. If we take $\sigma_i(d_i) > d_i$, by definition of the CEA rule we obtain

$$CEA_i(\sigma_1(d_1), ..., d_i, ..., \sigma_k(d_k)) = CEA_i(\sigma_1(d_1), ..., \sigma_i(d_i), ..., \sigma_k(d_k)),$$

therefore,

$$\pi_i(CEA_i(d_i; \sigma_{-i}(d_{-i})) | d_i) = \pi_i(CEA_i(\sigma_i(d_i); \sigma_{-i}(d_{-i})) | d_i).$$

$a.1.2$. If we consider $\sigma_i(d_i)$ such that $CEA_i(\sigma_1(d_1), ..., d_i, ..., \sigma_k(d_k)) \leq \sigma_i(d_i) \leq d_i$, then by the definition of the $CEA$ rule we are in the same situation as in ($a.1.1$).

$a.1.3$. If $\sigma_i(d_i) < CEA_i(\sigma_1(d_1), ..., d_i, ..., \sigma_k(d_k))$, then, independently of the relation between $\sum_{j=1}^{k} \sigma_j(d_j)$ and $r$, we will have that

$$CEA_i(\sigma_1(d_1), ..., \sigma_i(d_i), ..., \sigma_k(d_k)) < CEA_i(\sigma_1(d_1), ..., d_i, ..., \sigma_k(d_k)) < d_i.$$

Let us suppose that

$$\pi_i(CEA_i(d_i; \sigma_{-i}(d_{-i})) | d_i) < \pi_i(CEA_i(\sigma_i(d_i); \sigma_{-i}(d_{-i})) | d_i).$$

Now we take $0 < \alpha < 1$, such that

$$\alpha d_i + (1-\alpha) CEA_i(\sigma_1(d_1), ..., \sigma_i(d_i), ..., \sigma_k(d_k)) = CEA_i(\sigma_1(d_1), ..., d_i, ..., \sigma_k(d_k)).$$

Consider the optimal solutions $x^1$ and $x^2$ for problem (2) with $d_i$ and $CEA_i(\sigma_1(d_1), ..., \sigma_i(d_i), ..., \sigma_k(d_k))$, respectively. Thus, $\alpha x^1 + (1-\alpha) x^2$ is a feasible solution for problem (2) with $CEA_i(\sigma_1(d_1), ..., d_i, ..., \sigma_k(d_k))$. Therefore, we have that

$$\alpha value(S_i; d_i) + (1-\alpha) \pi_i(CEA_i(\sigma_i(d_i); \sigma_{-i}(d_{-i})) | d_i) \leq \pi_i(CEA_i(d_i; \sigma_{-i}(d_{-i})) | d_i),$$

but, by definition of $value(S_i; d_i)$, this should also be greater than $\pi_i(CEA_i(d_i; \sigma_{-i}(d_{-i})) | d_i)$ which is a contradiction.

$a.2$. $CEA_i(\sigma_1(d_1), ..., d_i, ..., \sigma_k(d_k)) = d_i$. Then two cases can arise:

$a.2.1$. $\sum_{j=1}^{k} \sigma_j(d_j) \leq r$. In this situation, we distinguish two subcases:

$a.2.1.1.$ If $\sigma_i(d_i) > d_i$, by definition of the $CEA$ rule we have

$$CEA_i(\sigma_1(d_1), ..., d_i, ..., \sigma_k(d_k)) < CEA_i(\sigma_1(d_1), ..., \sigma_i(d_i), ..., \sigma_k(d_k)).$$

But by definition of $d_i$, we know that

$$\pi_i(CEA_i(d_i; \sigma_{-i}(d_{-i})) | d_i) = value(S_i; d_i) \geq \pi_i(CEA_i(\sigma_i(d_i); \sigma_{-i}(d_{-i})) | d_i).$$

$a.2.1.2.$ If $\sigma_i(d_i) < d_i$, by definition of the $CEA$ rule we have

$$CEA_i(\sigma_1(d_1), ..., d_i, ..., \sigma_k(d_k)) > CEA_i(\sigma_1(d_1), ..., \sigma_i(d_i), ..., \sigma_k(d_k)),$$

and using the definition of $d_i$, we obtain that

$$\pi_i(CEA_i(d_i; \sigma_{-i}(d_{-i})) | d_i) = value(S_i; d_i) \geq \pi_i(CEA_i(\sigma_i(d_i); \sigma_{-i}(d_{-i})) | d_i).$$

$a.2.2.$ $\sum_{j=1}^{k} \sigma_j(d_j) > r$. Now, we distinguish two subcases:

$a.2.2.1.$ If $\sigma_i(d_i) > d_i$, using the definition of the $CEA$ rule we can derive

$$CEA_i(\sigma_1(d_1), ..., d_i, ..., \sigma_k(d_k)) \leq CEA_i(\sigma_1(d_1), ..., \sigma_i(d_i), ..., \sigma_k(d_k)).$$

But by definition of $d_i$, we know that

$$\pi_i(CEA_i(d_i; \sigma_{-i}(d_{-i})) | d_i) = value(S_i; d_i) \geq \pi_i(CEA_i(\sigma_i(d_i); \sigma_{-i}(d_{-i})) | d_i).$$

$a.2.2.2.$ If we consider $\sigma_i(d_i) < d_i$, by definition of the CEA rule we have

$$CEA_i(\sigma_1(d_1), ..., d_i, ..., \sigma_k(d_k)) > CEA_i(\sigma_1(d_1), ..., \sigma_i(d_i), ..., \sigma_k(d_k)),$$

and taking into account the definition of $d_i$, we obtain

$$\pi_i(CEA_i(d_i; \sigma_{-i}(d_{-i})) | d_i) = value(S_i; d_i) \geq \pi_i(CEA_i(\sigma_i(d_i); \sigma_{-i}(d_{-i})) | d_i).$$

Therefore, $\sigma_i(d_i) = d_i$ is a dominant strategy for $i$. Thus, $\sigma_i(d_i) = d_i$, for all $i$, is a dominant strategy equilibrium and the $CEA$ rule is incentive compatible. ■